\documentclass[aps,prd,preprintnumbers,superscriptaddress,
showkeys,showpacs,byrevtex,fleqn]{revtex4}
\usepackage{amsmath,amsfonts,amssymb,amscd,amsxtra,amsthm}
\usepackage{graphicx}
\usepackage{epstopdf}
\usepackage{bm}

\usepackage[normalem]{ulem} 
\usepackage[dvipsnames]{xcolor} 
\renewcommand\sout{\bgroup \color{red} \ULdepth=-.5ex \ULset}

\begin{document}
\preprint{INHA-NTG-01/2015}
\preprint{PKNU-NuHaTh-2015-01}
\title{Weak $K\to \pi$ generalized form factors and transverse
  transition quark-spin density from the instanton vacuum}
\author{Hyeon-Dong Son}
\email[E-mail: ]{hdson@inha.edu}
\affiliation{Department of Physics, Inha University, Incheon 402-751,
Republic of Korea}
\author{Seung-il Nam}
\email[E-mail: ]{sinam@pknu.ac.kr}
\affiliation{ Department of Physics, Pukyong National University,
Busan 608-737, Republic of Korea}
\author{Hyun-Chul Kim}
\email[E-mail: ]{hchkim@inha.ac.kr}
\affiliation{Department of Physics, Inha University, Incheon 402-751,
Republic of Korea}
\affiliation{School of Physics, Korea Institute for Advanced Study
  (KIAS),\\ Seoul 130-722, Republic of Korea}
\affiliation{Research Center for Nuclear Physics (RCNP),
Osaka University, Ibaraki, Osaka, 567-0047, Japan}
\date{\today}
\begin{abstract}
We investigate the generalized $K\to \pi$ transition vector and tensor
form factors, from which we derive the transverse quark spin density
in the course of the $K\to\pi$ transition, based on the nonlocal
chiral quark model from the instanton vacuum. The results of the
transition tensor form factor are in good agreement with recent data
of lattice QCD. The behavior of the transverse quark spin density of
the $K\to\pi$ transition turn out to be very similar to those of the
pion and the kaon.
\end{abstract}
\maketitle
\section{Introduction}
Semileptonic decay of the $K$ mesons ($K_{l3}$ decay)
provides a solid basis for testing various features of the Standard
Model (SM). In particular, the $K_{l3}$ decay can be used for
determining  the Cabibbo-Kobayashi-Maskawa (CKM)
matrix~\cite{Cabibbo:1963yz, Kobayashi:1973fv} precisely within the
Standard Model  (SM)~(see for example a recent
analysis~\cite{Antonelli:2010yf} and references therein). Since the
$W$-boson exchange in the SM governs the physics
of the $K_{l3}$ decay, the $K\to \pi$ vector transition elements have
been mainly considered to describe the $K_{l3}$ decay, while other
terms such as the tensor component were set aside.
Several experimental collaborations have searched for
possible nonzero values of the $K\to \pi$ tensor form factors but
found that the results turned out to be more or less consistent with
the SM prediction of the null value of the tensor form
factors~\cite{Akimenko:1991fv,  Ajinenko:2002ec, Ajinenko:2001iz,
  Ajinenko:2003uz, Lai:2004kb}. On the other hand, extensions beyond
the SM (BSM) with supersymmetry shed new light on the role
of the tensor operator in describing various weak decay processes of
the kaon~\cite{Gabbiani:1996hi, Martinelli:1998hz,
  Herczeg:2001vk, Cirigliano:2009wk, Grzadkowski:2010es} (see also
recent reviews~\cite{Cirigliano:2013lpa, Cirigliano:2013xha} and
references therein). These tensor operators arising from the BSM
extensions reveal new physics originating at the TeV scale, which may
be checked due to recent experimental progress in the near future. In
the meanwhile, lattice quantum chromodynamics (LQCD) can also test
the reliability of these operators. Very recently, Baum et
al.~\cite{Baum:2011rm} computed the matrix elements of the
electromagnetic operator $\bar{\psi} \sigma_{\mu\nu}\psi
F^{\mu\nu}$~\cite{Gabbiani:1996hi} between the pion and the kaon
within LQCD, which may be related to the CP-violating part of the
$K\to\pi l^+l^-$ semileptonic decays.

The tensor operator has another prominent place on the transversity of
hadrons~\cite{Ralston:1979ys, Cortes:1991ja, Jaffe:1991kp,
  Jaffe:1991ra,Barone:2001sp}.
While the transversity of hadrons provides us with essential
information on the quark spin structure of hadrons, it is very
difficult to be measured experimentally owing to its chiral-odd nature
and the absence of its direct probe. However, using semi-inclusive
deep inelastic scattering processes, Anselmino et al. were able to
extract the transverse parton distribution functions of the
nucleon and the corresponding tensor charges~\cite{Anselmino:2007fs,
  Anselmino:2008jk, Anselmino:2013vqa, Bacchetta:2012ty}. While the
transversity of the nucleon was extensively studied, the
transversities of the $\pi$ and $K$ mesons received little attention
again because of experimental difficulties to measure them. In the
meanwhile, it was found that the tensor form factors of hadrons can be
understood as generalized form factors that are defined as the Mellin
moments of generalized parton distributions (GPDs)~(see
reviews~\cite{Goeke:2001tz, Diehl:2003ny, Belitsky:2005qn} for
details). Moreover, the tensor form factors unveil the transverse
quark spin structure inside a hadron in the transverse
plane~\cite{Diehl:2005jf} with the proper probabilistic interpretation
of the transverse quark densities~\cite{Burkardt:2000za,
  Burkardt:2002hr}. Recently, QCDSF/UKQCD Collaborations
announced the first results for the pion transversity on
lattice~\cite{Brommel:2007xd}. They also presented the probability
density of the polarized quarks inside the pion, combining the
electromagnetic form factor of the pion~\cite{Brommel:2006ww} with its
tensor form factor. It was demonstrated in Ref.~\cite{Brommel:2007xd}
that when the quarks are transversely polarized, their spatial
distribution in the transverse plane is strongly distorted.
In addition, the $K\to \pi$ transitions can be also investigated from
a different point of view. Exclusive or semi-exclusive weak processes
may provide information on the $K\to \pi$ transitions via the weak
GPDs. In fact, the weak GPDs of baryons have been already examined in
Refs.~\cite{Psaker:2006gj, Kopeliovich:2012dr,
  Kopeliovich:2014pea}. It is thus worthwhile to
study the $K\to \pi$ transition GPDs. The $K\to \pi$ transition GPDs
provide much more information than the $K_{l3}$ form factors and the
tensor form factors, since they include all information about the
$K\to \pi$ transition generalized form factors, as mentioned above.

Thus, in the present work, we want to investigate the generalized
transition vector form factors and the multi-faceted generalized
tensor form factors in the context of the $K\to \pi$
transition. In Ref.~\cite{Nam:2007fx}, two of the authors have
investigated the transition vector form factors of the $K_{l3}$
decay, based on the low-energy effective chiral action from the
instanton vacuum. However, Ref.~\cite{Nam:2007fx} concentrated mainly
on the $K_{l3}$ decay. In this work, we extend the previous
investigation by computing the $K\to \pi$ transition vector and tensor
form factors also in the space-like region. Once we have these form
factors, we can immediately study the transverse quark spin densities
of the $K\to \pi$ transition.

In the present work, We want to utilize the nonlocal chiral quark
model (NL$\chi$QM) from the instanton vacuum to compute the
$K\to \pi$ transition vector and tensor form factors, aiming at
examining the transverse quark spin densities in the course of the
$K\to\pi$ transition. The NL$\chi$QM from the instanton was first
derived by Diakonov and Petrov~\cite{Diakonov:1985eg, Diakonov:2002fq}
in the chiral limit and was extended beyond the chiral
limit~\cite{Musakhanov:1998wp, Musakhanov:2002vu,
  Musakhanov:2002xa}. Since the instanton vacuum realizes the
spontaneous chiral symmetry breaking (S$\chi$SB) naturally via quark
zero modes, the NL$\chi$QM from the instanton vacuum provides a good
framework to study the vector and tensor form factors of the $K\to\pi$
transition. In fact, the model has been proven to be successful in
reproducing experimental data or in comparison with the results of
LQCD for the $\pi$ and $K$ mesons such as the low-energy constants of
the chiral Lagrangians~\cite{Franz:1999ik, Choi:2003cz},
electromagnetic form factors~\cite{Nam:2007gf}, meson distribution
amplitudes~\cite{Petrov:1998kg, Polyakov:1998td, Polyakov:1999gs,
  Nam:2006au, Nam:2006mb, Nam:2006sx}, semileptonic
decays~\cite{Nam:2007fx}, tensor form factors~\cite{Nam:2010pt,
  Nam:2011yw}, etc~\cite{Kim:2008in}.

The NL$\chi$QM is characterized by the two phenomenological
parameters, i.e., the average instanton size ($\bar{\rho}\approx1/3$
fm) and the average inter-instanton distance ($\bar{R}\approx1$ fm).
An essential advantage of this approach lies in the fact that  the
normalization point is naturally given by the average size of
instantons and is approximately equal to $\rho^{-1} \approx 0.6
\,\mathrm{GeV}$. This fact is essential, in particular, when one
calculates the matrix elements of the tensor current, since they are
scale-dependent. To compare the results of the tensor form factors
from any model, the normalization scale should be well defined such
that the results can be compared to those from other models or from
LQCD. The values of the $\bar{\rho}$̄ and $\bar{R}$ were estimated
many years ago phenomenologically~\cite{Shuryak:1981ff} as well as
theoretically~\cite{Diakonov:1983hh, Diakonov:1985eg}. Once the above-mentioned
two parameters $\bar{\rho}$̄ and $\bar{R}$̄ are determined, the
NL$\chi$QM from the instanton vacuum does not have any adjustable
parameter. Furthermore, this approach was supported by several LQCD
studies of the QCD vacuum~\cite{Chu:1994vi, Negele:1998ev,
  DeGrand:2001tm} and the momentum dependence of the dynamical quark
mass from the instanton vacuum~\cite{Diakonov:2002fq} is in a
remarkable agreement with those from LQCD~\cite{Faccioli:2003qz,
  Bowman:2004xi}.

The present work is organized as follows: In Section II, we
introduce the $K\to \pi$ transition GPDs based on which the
generalized form factors are defined. We also present the definition
of the transverse quark spin densities of the $K\to \pi$
transition. In Section III, we show how to compute the transition
vector and tensor form factors within the framework of the N$\chi$QM.
In Section IV, we present the results and discuss them. The final
section is devoted to the summary of the present work and discuss
future perspectives related to the transition GPDs and generalized
form factors.
\section{Generalized form factors and quark spin density of
the $K\to\pi$ transition}
The transition vector (tensor) GPDs $H^{K\pi} (x,
\xi,t)\,(E_{T}^{K\pi} (x, \xi, t))$ for the $K\to \pi$ transition are
defined respectively in terms of the matrix element of the vector
(tensor) nonlocal operators between the $K^0$ and the $\pi^-$ states:
\begin{eqnarray}
  \label{eq:wgpds}
2P^+ H^{K\pi} (x, \xi, t) &=& \int
\frac{d\lambda}{2\pi}    e^{ix\lambda (P \cdot n)}
\langle \pi^-(p') | \bar{s} ( - \lambda n /2)
\gamma^+ [-\lambda n/2,\lambda n/2] u(\lambda n/2) | K^0 (p)
\rangle, \cr
\frac{ P^+\Delta^{j}-\Delta^jP^+}{m_K}
E_{T}^{K\pi} (x, \xi, t) &=& \int \frac{d\lambda}{2\pi} e^{ix\lambda
  (P \cdot n)} \langle \pi^-(p') | \bar{s} ( - \lambda n /2)i
\sigma^{+j} [-\lambda n/2,\lambda n/2] u(\lambda n/2) | K^0 (p)
\rangle,
\end{eqnarray}
where $n$ denotes the light-like auxiliary vector. The momenta $p$ and
$p'$ correspond to those of the kaon and the pion, respectively. The $P$
represents the average momentum of the kaon and pion momenta $P^\mu
=(p^\mu + p'^\mu )/2$, whereas $\Delta$ corresponds to the momentum
transfer $\Delta^\mu = p'-p$, the square of which is expressed as
$t=\Delta^2$. $P^+=(P^0+P^3)/\sqrt{2}$ and $\Delta^j$ are expressed in
the light-cone basis. The index $j$ labels the transverse component,
i.e. $j=1$ or $j=2$. The kaon mass in the denominator is introduced to
define the tensor transition GPD $E_{T}^{K\pi} (x, \xi, t)$ to be
dimensionless. The gauge connection $[-\lambda n/2,\lambda
n/2]=P\exp[ig\int_{-\lambda n/2}^{\lambda n/2} dx^- A^+(x^-n_-)]$ can
be suppressed in the light cone gauge.
The generalized transition vector form factors $A^{K\pi}_{n+1,i+1}$
and $C^{K\pi}_{n+1,i+1}$ are defined by the following matrix elements:
\begin{align}
\label{eq:AG}
\langle\pi^-(k)|\mathcal{O}^{\mu\mu_1\cdots\mu_{n}}_V |K^0(p)\rangle
&=\mathcal{S}\left[
2 P^\mu P^{\mu_1} \cdots P^{\mu_n} A^{K\pi}_{n+1,0}(t)
+2 \sum^n_{i=1,\mathrm{odd}}\Delta^\mu \Delta^{\mu_1}
 \cdots \Delta^{\mu_i}P^{\mu_{i+1}} \cdots P^n A^{K\pi}_{n+1,i+1}(t)
 \right. \cr &+\left.
 2 \sum^n_{i=0,\mathrm{even}}\Delta^\mu \Delta^{\mu_1} \cdots \Delta^{\mu_i}
P^{\mu_{i+1}} \cdots P^n C^{K\pi}_{n+1,i+1}(t)
\right],
\end{align}
where the generalized vector transition operator is expressed as
\begin{equation}
\label{eq:OPv}
\mathcal{O}^{\mu\mu_1\cdots\mu_{n}}_V =
\mathcal{S}
\left[\bar{s} (\gamma^\mu i\tensor{D}^{\mu_1})
\cdots(i\tensor{D}^{\mu_{n}})u \right].
\end{equation}
The operation $\mathcal{S}$ means the symmetrization in $
(\mu,\cdots,\mu_n)$ with the trace terms subtracted in all
indices. $D_\mu$ indicates the hermitized covariant derivative
$i\tensor{D}_\mu\equiv (i\roarrow{D}_\mu-i\loarrow{D}_\mu)/2$ in
QCD. Note that the $A_{1,0}$ and $C_{1,1}$ are related to the
form factors $f_{l+}=A_{1,0}^{K\pi}$ and $f_{l-}=2C_{1,1}^{K\pi}$ of the
$K_{l3}$ decay, which are defined as
\begin{equation}
  \label{eq:Kl3}
\langle \pi^-(p')|\bar{s}\gamma^{\mu}
u |K^0(p)\rangle
=2P^\mu f_{l+}(t) + \Delta^\mu f_{l-}(t),
\end{equation}
where $s$ and $u$ denote the strange and up quark fields.
The generalized transition tensor form factors $B_{T\,n,i}^{K\pi}$ can
be also defined by the following matrix element
\begin{equation}
\label{eq:BG}
\langle\pi^-(p')|\mathcal{O}^{\mu\nu\mu_1\cdots\mu_{n-1}}_T|K^0(p)\rangle
=\mathcal{AS}\left[\frac{(P^{\mu}\Delta^{\nu}-\Delta^{\mu}P^{\nu})}{m_{K}}
\sum^{n-1}_{i=0} \Delta^{\mu_1}\cdots \Delta^{\mu_i}
P^{\mu_{i+1}}\cdots P^{\mu_{n-1}}B_{T\,n,i}^{K\pi}(t)\right],
\end{equation}
where the generalized tensor transition operator is expressed as
\begin{equation}
\label{eq:OP}
\mathcal{O}^{\mu\nu\mu_1\cdots\mu_{n-1}}_T=
\mathcal{AS}
\left[\bar{s}\sigma^{\mu\nu}(i\tensor{D}^{\mu_1})
\cdots(i\tensor{D}^{\mu_{n-1}})u \right].
\end{equation}
The operations $\mathcal{A}$ and $\mathcal{S}$ mean the
anti-symmetrization in $(\mu,\nu)$ and symmetrization in
$(\nu,\cdots,\mu_{n-1})$ with the trace terms subtracted in all the
indices. The antisymmetric tensor is defined as $\sigma_{\mu\nu} =
i \left( \gamma_\mu \gamma_\nu - \gamma_\nu \gamma_\mu
\right)/2$.
Note that there are odd and even terms of $\xi$ in Eqs. (\ref{eq:OPv})
and (\ref{eq:OP}), respectively, due to the fact that the matrix elements
for the $K \to \pi$ transitions do not vanish under the time-reversal
transformation. The leading-order transition vector and tensor form
factors are then expressed as
\begin{eqnarray}
  \label{eq:TVFF}
\langle \pi^-(p') | \bar{s}\gamma_{\mu} u | K^0(p) \rangle
&=& 2 P_\mu A^{K\pi}_{1,0}(t) + 2\Delta_\mu C^{K\pi}_{1,1} (t),\\
\langle \pi^-(p') | \bar{s}\sigma_{\mu\nu} u | K^0(p) \rangle
&=& \left(  \frac{P_\mu \Delta_\nu - P_\nu \Delta_\mu}{m_K}
\right)B^{K\pi}_{T\,1,0}(t),\label{eq:TTFF}
\end{eqnarray}
in which we are mainly interested.
Combining Eqs.(\ref{eq:wgpds}, \ref{eq:AG}) with Eq. (\ref{eq:BG}),
we find the formula for $n$th order Mellin moments of the
vector and tensor transition GPDs:
\begin{eqnarray}
\label{eq:poly}
\int dx\,x^n H^{K\pi} (x, \xi, t) &=& A_{n+1,0}^{K\pi}(t) +
 \sum_{i=1,\mathrm{odd}}^n (-2\xi)^{i+1} A_{n+1, i+1}^{K\pi} (t)+
 \sum_{i=1,\mathrm{odd}}^{n+1} (-2\xi)^i C_{n+1, i}^{K\pi} (t)
 , \cr
\int dx\,x^n E_{T}^{K\pi} (x, \xi, t) &=& \sum_{i=0}^n
(-2\xi)^i B_{T\,n+1, i}^{K\pi} (t).
\end{eqnarray}
so that the transition vector and tensor form factors can be
identified respectively as the first moments of the vector and tensor
transition GPDs
\begin{equation}
\int dx\, H^{K\pi}(x,\xi,t)
= A_{1,0}^{K\pi} - 2\xi C_{1,1}^{K\pi}, \;\;\;\;
 \int dx\,E_{T}^{K\pi} (x, \xi, t) = B_{T\,1,0}^{K\pi}(t),
\end{equation}
where the skewedness parameter is defined as $\xi =
-\Delta^+/(2P^+)$.
Finally the spin distribution of the transversely polarized quark
in the course of the $K\to \pi$ transition is written
as follows~\cite{Brommel:2007xd}:
\begin{equation}
\label{eq:DENSITY}
\rho^{K\pi}_{1}(b,\bm s_{\perp})=
\frac{1}{2}\left[A_{1,0}^{K\pi}(b^{2})
-\frac{s^{i}_{\perp}\epsilon^{ij}b^{j}}{m_{K}}
\frac{\partial B^{K\pi}_{T1,0}(b^{2})}{\partial b^{2}}\right],
\end{equation}
with the Fourier transformations of the transition vector and tensor
form factors
\begin{equation}
\label{eq:FT}
\mathcal{F}_{1,0}^{K\pi}(b_\perp)=\frac{1}{(2\pi)^2}\int d^{2}
\Delta^{-i{\bm b_\perp}\cdot {\bm \Delta}}
\mathcal{F}_{1,0}^{K\pi}(t)=\frac{1}{2\pi}\int^\infty_0
Q\,dQ\,J_0(bQ)\,\mathcal{F}_{1,0}^{K\pi}(Q^{2}).
\end{equation}
The form factors $\mathcal{F}_{1,0}^{K\pi}(t)$ and densities
$\mathcal{F}_{1,0}^{K\pi}(b_\perp)$ stand generically either for the
transition vector ones or the tensor ones.
The $\bm s_\perp=(s_x,s_y)$ stands for the fixed transverse spin
of the quark. We choose the $z$ direction for the quark longitudinal
momentum for simplicity and select the $x$ axis in the transverse
plane for the quantization of the spin of the quark in the course of the
$K\to \pi$ transition in the transverse plane, that is,
$s_{\perp}=(\pm1,0)$.
\section{Nonlocal chiral quark model from the
instanton vacuum}
The NL$\chi$QM can be derived from the instanton
liquid model for the QCD vacuum.  Starting from the QCD partition
function in the one-loop approximation, in which the classical
background field (instantons) and  one
can express the following partition function~\cite{Callan:1977gz,
  Diakonov:1983hh, Diakonov:1985eg}
\begin{equation}
Z_{\mathrm{reg},\mathrm{norm}}^{\mathrm{1-loop}} \;=\; \frac{1}{N_+!
  N_-!} \int \prod_I^{N_++N_-} d\xi_I d_0(\rho_I)
\exp(-U_{\mathrm{int}}) \mathrm{Det}[m,\,M_{\mathrm{cut}}],
\end{equation}
where $N_+$ and $N_-$ denote the number of instantons and
anti-instantons, respectively. $\xi_I$ designates generically the
collective coordinates including the center positions of the
instantons $z_I$, their sizes $\rho_I$, and orientations of the
instantons expressed in terms of $\mathrm{SU}(N_c)$ matrices in the
adjoint representation. $d_0(\rho_I)$ represents the one-instanton
weight, which was originally derived by 't Hooft in
SU(2)~\cite{'tHooft:1976fv} and
by Bernard in SU(3) and $\mathrm{SU}(N_c)$~\cite{Bernard:1979qt} in
the one-loop approximation:
\begin{equation}
  \label{eq:d_0}
d_0(\rho_I) = \frac{C_{N_c}}{\rho_I^5} \beta(M_{\mathrm{cut}})^{2N_c}
\exp[-\beta(\rho_I)],
\end{equation}
where $\beta(\rho_I)$ is the inverse of the strong coupling constant
in the one-loop approximation
\begin{equation}
\beta(\rho_I) = \frac{8\pi^2}{g^2(\rho_I)} = b
\log\left(\frac1{\Lambda_{\mathrm{PV}} \rho_I}\right)
\label{eq:beta}
\end{equation}
with $b=11 N_c/3-2N_f/3$. The QCD scale parameter
$\Lambda_{\mathrm{PV}}$ here is given in the Pauli-Villars
regularization and is related to that in the $\overline{\mathrm{MS}}$
scheme $\Lambda_{\mathrm{PV}}=1.09\Lambda_{\overline{\mathrm{MS}}}$.
The coefficient $C_{N_c}$ depends on renormalization schemes and is
given in the Pauli-Villars scheme as
\begin{equation}
C_{N_c} =\frac{4.66 \exp(-1.68 N_c)}{\pi^2 (N_c-1)!(N_c-2)!}.
\end{equation}
The effective instanton size distribution which is related to
$d_0(\rho_I)$ is reduced to a $\delta$-function in the large $N_c$
limit because of the presence of $b$ in Eq.(\ref{eq:beta}), which
picks up the average size of the instanton $\bar{\rho}$.
The instanton interaction potential $U_{\mathrm{int}}$ was derived and
studied in Ref.~\cite{Diakonov:1983hh}. The regularized and normalized
fermionic determinant $\mathrm{Det}$ depends on the Pauli-Villars
cut-off mass $M_{\mathrm{cut}}$.

Since we aim at deriving the $K\to\pi$ transition form factors in the
present work, we need to include the external sources for the
vector and tensor fields in the fermionic determinant
$\widetilde{\mathrm{Det}}$, which is given as a functional of
$V_\mu$ and $T_{\mu\nu}$~\cite{Kim:2004hd}:
\begin{equation}
\label{eq:Det}
\widetilde{\mathrm{Det}} :=
\mathrm{Det}(i\rlap{/}{\partial} + g \rlap{\,/}{A} +
\rlap{/}{V} + \sigma_{\mu\nu} T_{\mu\nu} + i \hat{m}),
\end{equation}
where $A_\mu$ is the gluon field with the gauge coupling constant $g$
and $\hat{m} = \mathrm{diag}(m_u,m_d,m_s)$ denotes the current quark
mass matrix that shows explicit chiral and flavor SU(3) symmetry
breaking, of which their numerical values are given as $m_u= m_d=
5\,\mathrm{MeV}$ and $m_s = 150\,\mathrm{MeV}$. The fermionic
determinant $\widetilde{\mathrm{Det}}$ can be divided into two parts
corresponding to the low and high Dirac eigen-frequencies with respect
to an arbitrary splitting parameter $M_1$:
$\widetilde{\mathrm{Det}}(m,M_{\mathrm{cut}}) :=
\widetilde{\mathrm{Det}}_{\mathrm{low}}(m,M_1)\,
\widetilde{\mathrm{Det}}_{\mathrm{high}}(M_1,M_{\mathrm{cut}})$.
The high-frequency part
$\widetilde{\mathrm{Det}}_{\mathrm{high}}$ was shown to contribute to
the statistical weights of individual instantons. That is, it
influences mainly the renormalization of the coupling constant in a
sense of the renormalization group equation. On the other hand,
The low-frequency part $\widetilde{\mathrm{Det}}_{\mathrm{low}}$
can only be treated approximately, the would-be zero modes being only
taken into account. It was proven that the
$\widetilde{\mathrm{Det}}_{\mathrm{low}}$ depends weakly on the scale
$M_1$ in a broad range of $M_1$, so that the
the matching between $\widetilde{\mathrm{Det}}_{\mathrm{high}}$ and
$\widetilde{\mathrm{Det}}_{\mathrm{low}}$ turns out to be
smooth~\cite{Diakonov:1985eg}.
The natural choice of the parameter $M_1$ can be taken to be roughly
$M_1\sim 1/\bar{\rho}$, where $\bar{\rho}$ is the average size
of instantons $1/\bar{\rho}\simeq 600\, \mathrm{MeV}$. Thus, as
mentioned already, $1/\bar{\rho}$ can be considered as the natural
scale of the present model. Of course the choice of $\bar{\rho}\simeq
600$ MeV is not strict but has some ambiguity. We will discuss this
ambiguity in the context of the tensor form factor later.

The low-frequency part
$\widetilde{\mathrm{Det}}_{\mathrm{low}}$ was derived in
Refs.~\cite{Diakonov:1985eg,Diakonov:1995qy,Kim:2004hd} and its
explicit form is written as
\begin{eqnarray}
\label{part-func}
\widetilde{\mathrm{Det}}_{\mathrm{low}} &=&
 \left(\det(i\rlap{/}{\partial} +\rlap{/}{V} +\sigma \cdot T +
 i\hat{m})\right)^{-1} \int \prod_{f}D\psi_f D\psi^{\dagger}_{f}
\\\nonumber
&\times&
 \exp{\left(\int d^4 x
\psi_{f}^{\dagger} (i\rlap{/}{\partial} \,+\,\rlap{/}{V}
 \,+\sigma \cdot T\, +\, im_f )\psi_{f}\right)}
 \prod_{f}\left\{\prod_{+}^{N_{+}}
V_{+,f}[\psi_{f}^{\dagger},\psi_f ]
\prod_{-}^{N_{-}}V_{-,f}[\psi_{f}^{\dagger},\psi_f ]\right\}\; ,
\end{eqnarray}
where
\begin{equation}
  \label{eq:FermionVertex}
\tilde V_{\pm,f}[\psi_{f}^{\dagger},\psi_f ]=\int d^4 x
\left(\psi_{f}^{\dagger} (x)\,L_{f}(x,z) i\rlap{/}{\partial}
\Phi_{\pm, 0} (x; \xi_{\pm})\right) \int d^4 y \left(\Phi_{\pm ,
0} ^\dagger (y; \xi_{\pm} ) (i\rlap{/}{\partial} L^{+}_{f}(y,z)
\psi_{f} (y)\right).
\end{equation}
The $\psi_f$ denotes the quark field, given flavor $f$. The $m_f$ is
the current quark mass corresponding to $\psi_f$. The $N_+$ and $N_-$
stand for the number of instantons and anti-instantons. The gauge
connection $L_f$ is defined as
\begin{equation}
  \label{eq:GaugeConnection}
L_f(x,z) := \mathrm{P}\exp\left(\int_z^x d \zeta_\mu
  V_\mu(\zeta)\right),
\end{equation}
which is essential to make the nonlocal effective action
gauge-invariant and should be attached to each fermionic line. The
$\Phi_{\pm,0} (x; \xi_{\pm})$ represents the zero-mode solution of the
Dirac equation in the instanton ($A_{\mu,+}$) and anti-instanton
($A_{\mu,-}$) fields
$(i\rlap{/}{\partial}+\rlap{\hspace{1.5pt}/}{A}_{\pm})\Phi_{\pm,0}
(x;\xi_{\pm}) = \lambda_n \Phi_{\pm,0} (x;\xi_{\pm})$.
Having exponentiated and bosonized the fermionic interactions
$V_{\pm,f}$, and having averaged the low-frequency part of the
fermionic determinant $\widetilde{\mathrm{Det}}_{\mathrm{low}}$ over
collective coordinates $\xi_\pm$, we arrive at the effective chiral
partition function of which the detailed derivation can be found in
Refs.~\cite{Diakonov:1985eg,Diakonov:1995qy,
  Musakhanov:2001pc,Kim:2004hd}.

Since our main concern is to compute the $K\to \pi$ tensor generalized
form factors in the present work, we set the stage for them by
using the relevant effective chiral action of the NL$\chi$QM with the
external tensor source field $T_{\mu\nu}$ derived from
Eq.(\ref{part-func}): 
\begin{equation}
\label{ECA}
\mathcal{S}_{\mathrm{eff}}[T] =
- \mathrm{Sp} \mathrm{ln}
\left[ i \rlap{/}{\partial}+ i \hat{m} + i \sqrt{M}U^{\gamma_5}\sqrt{M}
+ T_{\mu\nu} \sigma_{\mu\nu}\right].
\end{equation}
Here, the functional trace $\mathrm{Sp}$ runs over the
space-time, color, flavor, and spin spaces. Note that isospin symmetry
is assumed. The nonlinear pseudo-Nambu-Goldstone boson field is
written as
\begin{equation}
\label{eq:NLGB}
U^{\gamma_5} = \exp \left(\frac{i \gamma_5}{F_\phi}\lambda\cdot\phi
\right),\,\,\,\, \phi=(\pi,K,\eta),
\end{equation}
where the pion and kaon weak-decay constants are chosen to be
$(F_\pi,F_K) = (93,\,113)$ MeV empirically. The pseudoscalar meson
fields are defined by
\begin{equation}
\label{eq:PGBF}
\lambda \cdot \phi = \sqrt{2} \left( \begin{array}{cccc}
\frac{1}{\sqrt{2}}\pi^0 +\frac{1}{\sqrt{6}} \eta & \pi^+ & K^+ \\
\pi^- & -\frac{1}{\sqrt{2}}\pi^0 +\frac{1}{\sqrt{6}}\eta & K^0 \\
K^- & \bar{K}^0 & -\frac{2}{\sqrt{6}}\eta \end{array}\right).
\end{equation}
For the numerical calculations, we use the mass values for the pion
and kaon as $(m_\pi,m_K)=(140,495)$ MeV throughout the present work,
taking the flavor SU(3) symmetry breaking into account. The
momentum-dependent dynamical quark mass, which is induced from the
nontrivial quark-instanton interactions and indicates SB$\chi$S, is
given by 
\begin{equation}
\label{eq:DQM}
M_f(k) = M_0 F^2(k) \left[ \sqrt{1+\frac{m_f^2}{d^2}} -
  \frac{m_f}{d}\right],
\end{equation}
where $M_0$ is the constituent quark mass at zero quark virtuality,
and is determined by the saddle-point equation, resulting in about
$350$ MeV~\cite{Diakonov:1985eg,Diakonov:2002fq}. The
form factor $F(k)$ arises from the Fourier transform of the quark
zero-mode solution for the Dirac equation with the instanton and has
the following form:
\begin{equation}
\label{eq:ff_dqm}
F(k) = 2\tau\left[ I_0 (\tau) K_1(\tau) -I_1 (\tau)K_0(\tau)
  -\frac{1}{\tau}I_1(\tau)K_1(\tau) \right],
\end{equation}
where $\tau\equiv\frac{|k|\bar{\rho}}{2}$. In this work, however, we
use the following parametrization for numerical convenience:
\begin{equation}
\label{eq:ff_dipole}
F(k)= \frac{2 \mu^2}{2 \mu^2 + k^2},
\end{equation}
where $\mu=1/\bar{\rho}=600\,\mathrm{MeV}$ can be regarded as the
renormalization scale of the model. In order to take into account the
explicit flavor SU(3) symmetry breaking effects properly, we modify
the dynamical quark mass with the $m_f$-dependent term given in the
bracket in the right-hand side of
Eq.~(\ref{eq:DQM})~\cite{Pobylitsa:1989uq,Musakhanov:2001pc} in such a
way that the instanton-number density $N/V$ is independent of the
current-quark mass, where $N$ and $V$ denote the number of instantons
and the four-dimensional volume, respectively. Pobylitsa
took into account the sum of all planar diagrams in expanding the quark
propagator in the instanton background in the large $N_c$
limit~\cite{Pobylitsa:1989uq}. Taking the limit of $N/(VN_c)\to 0$
leads to the term in the bracket of Eq.~(\ref{eq:DQM}). The parameter
$d$ is chosen to be $0.193$ GeV. It is worth noting that this
modification gives a correct hierarchy of the strengths for the chiral
condensates: $\langle\bar{u}u\rangle\approx\langle\bar{d}d\rangle>
\langle\bar{s}s\rangle$~\cite{Nam:2006ng}.

\begin{figure}[htp]
\includegraphics[width=10cm]{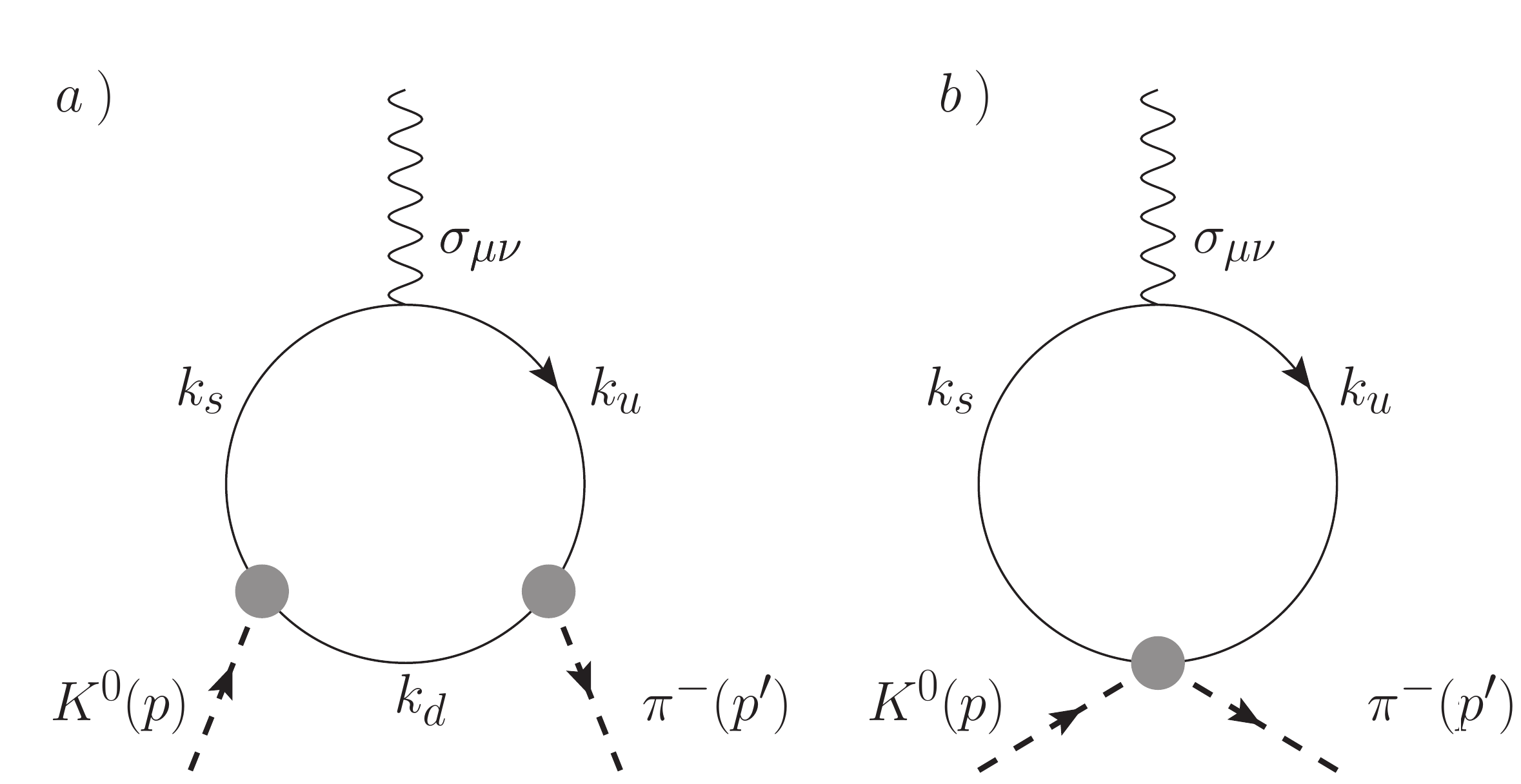}
\caption{Relevant Feynman diagrams for the $K\to\pi$ transition tensor
  form factor. The solid, dash, and wavy lines denote the quark,
  the pseudoscalar meson, and the tensor operator, respectively. The
  four momenta of the quarks are defined and explicitly given in
  Eq.~(\ref{eq:QLM}).}
\label{FIG1}
\end{figure}
The matrix element in Eq.~(\ref{eq:TTFF}) can be
straightforwardly derived by taking the functional derivative of
Eq.(\ref{ECA}) with respect to the pion, kaon, and external tensor
fields, resulting in
\begin{equation}
\label{eq:MATELE}
\langle \pi^-(p') | \bar{s}\sigma_{\mu\nu} u | K^0(p) \rangle
= -\frac{8N_c}{F_\pi F_K} \int \frac{d^4l}{(2\pi)^4}\left[
\frac{  \sqrt{M^2_{d}M_{u}M_{s}}  } { G_{u} G_{d} G_{s} }
\epsilon^{ijk}k_{i\mu}k_{j\nu} \bar{M}_{k f_k}
-\frac{ \sqrt{M_{u}M_{s}} } {2 G_{u} G_{s} }
\left( k_{s\mu} k_{u\nu} - k_{s\nu} k_{u\mu} \right)\right],
\end{equation}
where we introduced $\bar{M}_{f}(k^2_f)= m_{f} + M_{f}(k^2_f)$ and
$G_{f}= k_f^2 + \bar{M}_{f}^2$ with $f=(u,d,s)$. The first and second
terms inside the squared bracket in the right-handed side of
Eq.~(\ref{eq:MATELE}) correspond to the diagrams ($a$) and ($b$) of
Fig.~\ref{FIG1}, respectively. The quark four momenta shown in the
figure are defined as follows:
\begin{equation}
\label{eq:QLM}
 k_u = l + \frac{p}{2} + \frac{\Delta}{2},\,\,\,
 k_d = l - \frac{p}{2} - \frac{\Delta}{2},\,\,\,
 k_s = l + \frac{p}{2} - \frac{\Delta}{2}.
 \end{equation}
The four-momenta of the kaon at rest and the pion are defined
in the center-of-mass frame as
\begin{equation}
\label{frame2}
 p = \left(0, 0, 0, i E_K\right),\,\,\,\,
 p' = \left(- \sqrt{\left(\frac{t+m_K^2 +m_{\pi}^2}{2m_K}\right)
     -m_{\pi}^2}, 0, 0, i E_\pi \right).
\end{equation}

In order to compare our numerical results of the transition tensor
form factor with those of other works, it is crucial to know the
renormalization scale, since the tensor current is not the conserved
one. Results at two different scales are related by the following the
next-to-leading (NLO) order evolution equation~\cite{Gluck:1994uf,
  Barone:2001sp, Becirevic:2000zi}:
\begin{equation}
  \label{eq:RG}
B^{K\pi}_{T\,1,0}(\mu^2) =
\left(\frac{\alpha_s(\mu^2)}{\alpha_s({\mu_i^2})}\right)^{4/27}
\left[ 1-\frac{337}{486\pi}(\alpha_s(\mu_i^2)-\alpha_s(\mu^2))
\right]B^{K\pi}_{T\,1,0}(\mu_1)
\end{equation}
with the NLO strong coupling constant
\begin{equation}
\alpha_s^{\mathrm{NLO}} (\mu^2)=
\frac{4\pi}{9\ln(\mu^2/\Lambda_{\mathrm{QCD}}^2)}
\left[
1-\frac{64}{81}\frac{\ln\ln(\mu^2/\Lambda_{\mathrm{QCD}}^2)}{
  \ln(\mu^2/\Lambda_{\mathrm{QCD}}^2)}
\right].
\end{equation}
where $\mu_i$ denotes the initial renormalization scale, and
we take $N_f=3$ in the present work. Note that the scale dependence of
the tensor form factor given in Eq.(\ref{eq:RG}) is rather mild. As
will be shown explicitly in the next Section, the tensor form factor
is changed approximately by $10\,\%$ when one scales down from
$\mu=2\,\mathrm{GeV}$ to $\mu=0.6\,\mathrm{GeV}$. It 
indicates that even though we choose some higher or lower value of the
scale of the model, the result is not much changed. Thus, the
ambiguity in choosing the scale of the present model will have only a
tiny effect on the result of the tensor form factor when one scales it
to another normalization point.

\section{Numerical results and Discussions}
In this Section, we present the numerical results and discuss them.
We start with the $K\to \pi$ transition vector form factors.
While the kinematically accesible region for the $K\to \pi$
semileptonic form factors $f_{l+}$ and $f_{l-}$ is restricted to
$m_l^2\le t \le (m_K-m_\pi)^2$, where $m_l$ is the lepton mass
involved in the decay, the generalized transition vector form
factors $A_{1,0}^{K\pi}$ and $C_{1,1}^{K\pi}$ related to the
transition GPDs can be also defined in the spacelike region, since
they can be extracted in principle from exclusive weak processes.
\begin{figure}[htp]
\includegraphics[width=8.5cm]{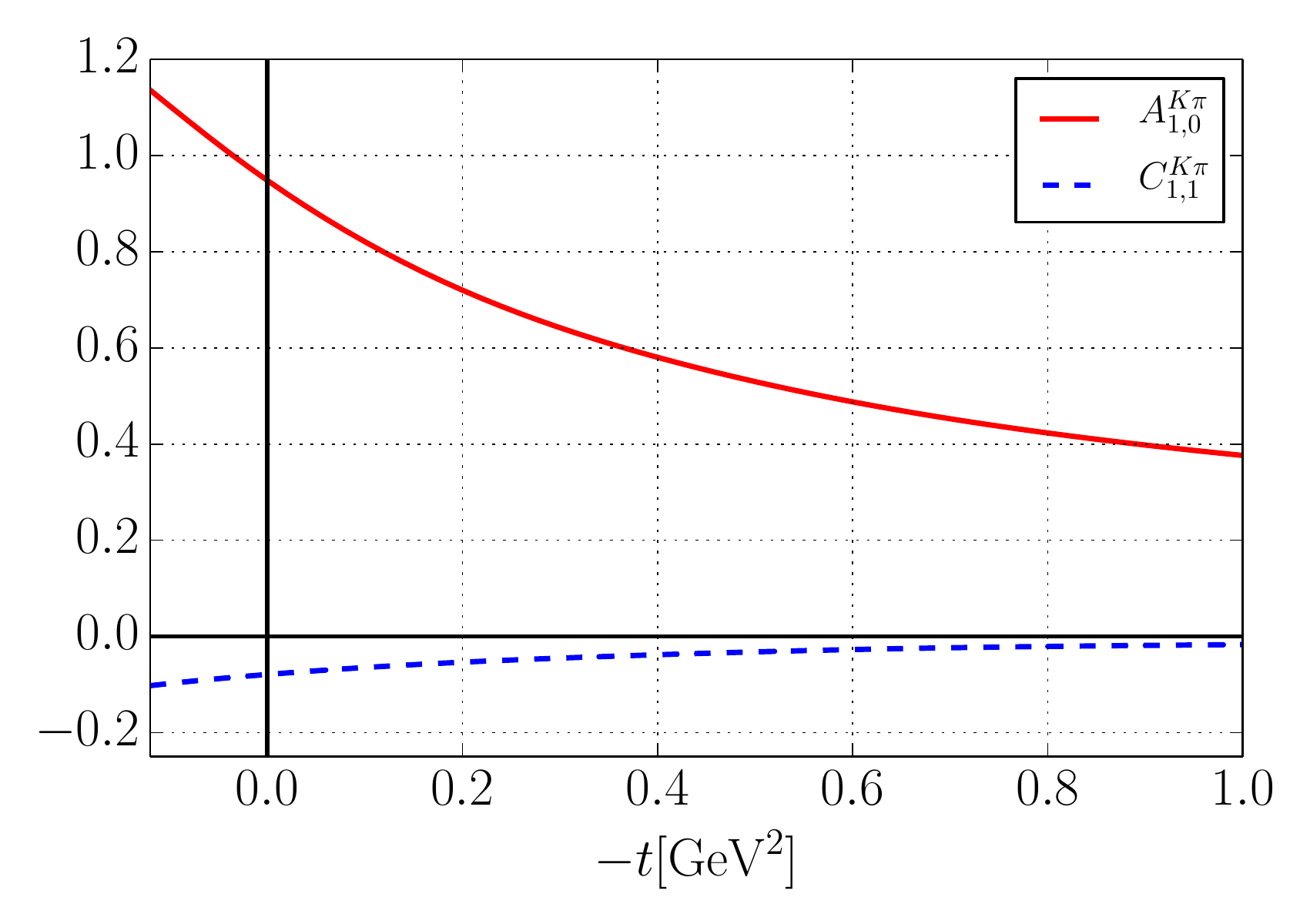}
\caption{(Color online) The $K\to \pi$ transition vector form factors
as functions of $t$ in the space-like region. The solid curve draws
the result of $A_{1,0}^{K\pi}$, whereas the dashed one depicts that of
$C_{1,1}^{K\pi}$.}
\label{fig:2}
\end{figure}
In Fig.~\ref{fig:2}, we show the results of the $K\to \pi$ transition
vector form factors $A_{1,0}^{K\pi}$ and $C_{1,1}^{K\pi}$ in the
space-like region. Both form factors fall off as $|t|$
increases. The magnitude of $A_{1,0}^{K\pi}$ turns out to be much
larger than that of $C_{1,1}^{K\pi}$. This can be understood from the
results for the $K\to \pi$ semileptonic decay~\cite{Nam:2007fx} in
which the magnitude of $f_{l+}(m_l^2)$ is approximately eight times
larger than that of $f_{l-}(m_l^2)$. It is the general tendency also
known from other approaches.

\begin{figure}[htp]
\begin{tabular}{cc}
\includegraphics[width=8.5cm]{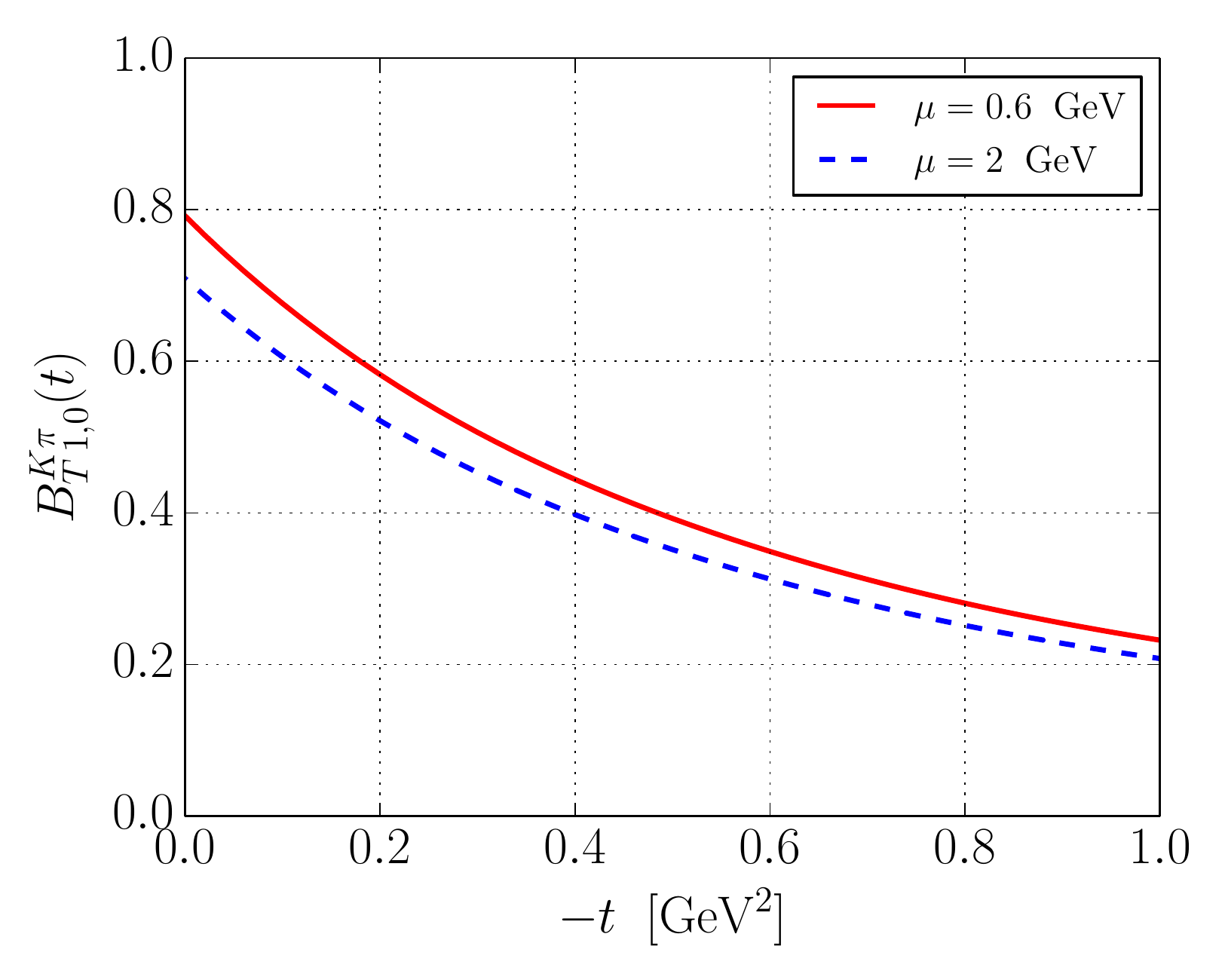}
\includegraphics[width=8.5cm]{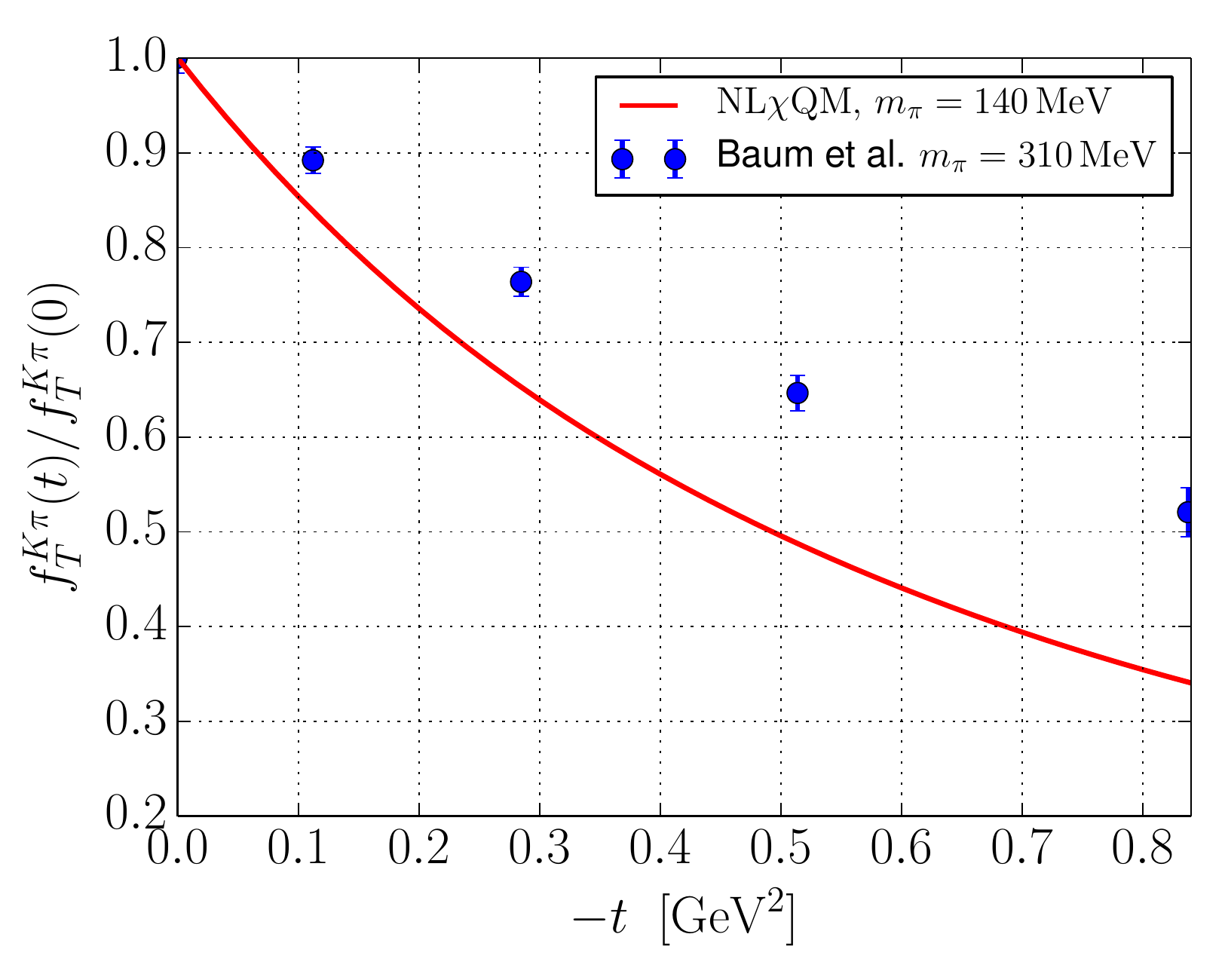}
\end{tabular}
\caption{(Color online) The$K\to \pi$ transition tensor form factors
  as functions of $-t$ in the space-like region. In the left panel,
  the solid curve draws the result of $B_{T1,0}^{K\pi}$ at $\mu=0.6$
  GeV, whereas the dashed one depicts that of  $B_{T1,0}^{K\pi}$ at
  $\mu=2.0$ GeV. The right panel compares the result of
  $B_{T1,0}^{K\pi}$ with that from lattice QCD at $\mu=2.0$ GeV.}
\label{fig:3}
\end{figure}
In the left panel of Fig.~\ref{fig:3}, we depict the transition tensor
form factors $B_{1,0}^{K\pi}$ as a function of $-t$ at two different
scales. Since it depends on the renormalization scale, we examine the
scale dependence of the transition tensor form factor, based on
Eq.(\ref{eq:RG}). The solid curve draws
the present result, which is given at the renormalization scale
$\mu=0.6$ GeV of the NL$\chi$QM, whereas the dashed one represents the
form factor at $\mu=2.0$ GeV, which corresponds to the scale of
LQCD~\cite{Baum:2011rm}. We observe that the transition tensor form
factor depends mildly on $\mu$. The value of the form factor at $t=0$
is given respectively as $B^{K^0\pi^-}_T(0)=0.792$ at $\mu=0.6$ GeV
and $B^{K^0\pi^-}_T(0)=0.709$ at $\mu=2$ GeV. That is, the magnitude
of the form factor is approximately reduced by $10\;\%$, when $\mu$ is
scaled up to $\mu=2$ GeV from $\mu= 0.6$ GeV. Since the scale factor
is an overall one, the $t$-dependence of the form factor is not
affected by the scaling. The right panel of Fig.~\ref{fig:3} draws the
transition tensor form factor normalized by its value at $t=0$ in
comparison with that of LQCD~\cite{Baum:2011rm} at $\mu=2$ GeV.
Note that  Ref.~\cite{Baum:2011rm} computed the transition tensor form
factor $f_T^{K\pi}(t)$ defined as
\begin{equation}
\label{eq:matrix_neutral}
\langle \pi^0(p') | \bar{s}\sigma_{\mu\nu} d | K^0(p) \rangle
=\left(p'_\mu p_\nu - p'_\nu p_\mu \right)
\frac{\sqrt{2}f^{K\pi}_{T}(t)}{m_K + m_\pi},
\end{equation}
which can be written in terms of $B_{T1,0}^{K\pi}$:
\begin{equation}
\label{eq:FFREL}
 f^{K\pi}_T(t) = \frac{m_K + m_\pi}{2m_K} B^{K\pi}_{T\,1,0}(t).
\end{equation}
At $t=0$, the value of the form factor $f^{K\pi}_T(0)$ is obtained to
be $f^{K\pi}_T(0) = 0.45$ at $\mu=2\,\mathrm{GeV}$, while the lattice
result becomes $f_T^{K\pi}(0) =
0.417\pm0.014(\mathrm{stat})\pm 0.05(\mathrm{syst})$ at the
physical pion mass after the extrapolation from $m_\pi=270$
MeV. Hence, the present result is in good agreement with the lattice
one. The present result of the form factor falls off faster than that
of LQCD. The reason can be found in the fact that the pion mass
employed in LQCD is still larger than the physical one. A similar
feature is found in the case of the nucleon tensor form
factor~\cite{Ledwig:2010tu, Ledwig:2010zq}. The lattice results of the
nucleon tensor form factors also fall off rather slowly.

Once we have derived the transition vector and tensor form factors, we
can proceed to the calculation of the transverse quark spin density in
the course of the $K\to \pi$ transition, using Eq.(\ref{eq:DENSITY}).
In doing so, it is more convenient to parameterize the form factors in
the $p$-pole type, which is usually employed in the lattice
calculation~\cite{Brommel:2007xd}:
\begin{equation}
\label{eq:PPFF}
\mathcal{F}^{K\pi}_{1,0}(t)=\mathcal{F}^{K\pi}_{1,0}(0)\left(
  1+\frac{t}{pM_p^2} \right)^{-p},
\end{equation}
so that the Fourier transform can be easily carried out.
Having fitted the results of the form factors shown in Fig.~\ref{fig:2}
and the left panel of Fig.~\ref{fig:3} , we are able to determine
the parameters as $p=0.850$
and $M_p = 1.312 \,\mathrm{GeV}$ for $A_{1,0}^{K\pi}$ and
$p=2.172$ and $M_p=0.776\,\mathrm{GeV}$ for $B_{T\,1,0}^{K\pi}$,
respectively, at $\mu=0.6$ GeV. Using these values, we can easily
derive the quark spin transverse density in the course of the
$K\to\pi$ transition, which is defined in Eq.(\ref{eq:DENSITY}).

\begin{figure}[htp]
\begin{tabular}{cc}
\includegraphics[width=8.5cm]{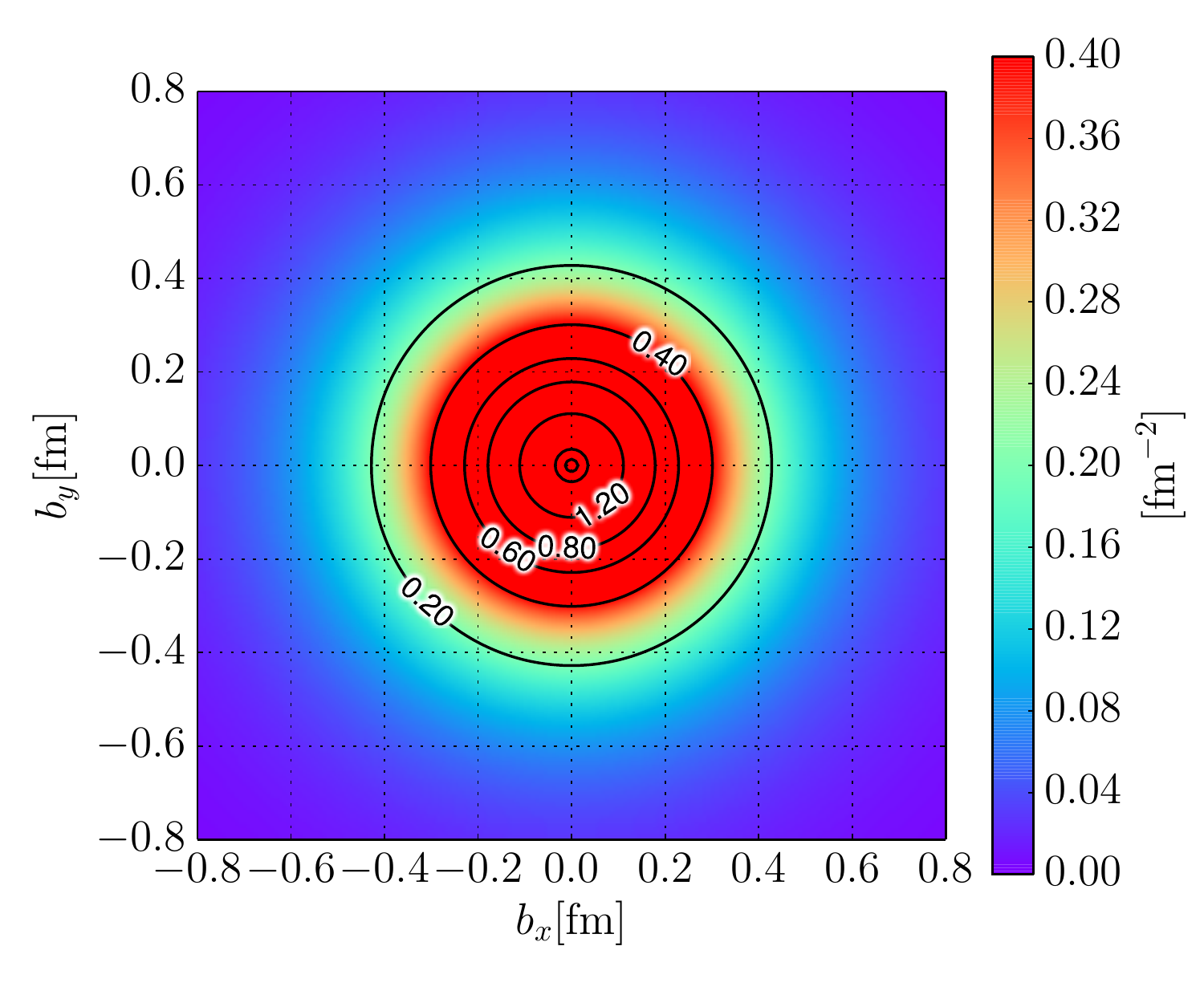}
\includegraphics[width=8.5cm]{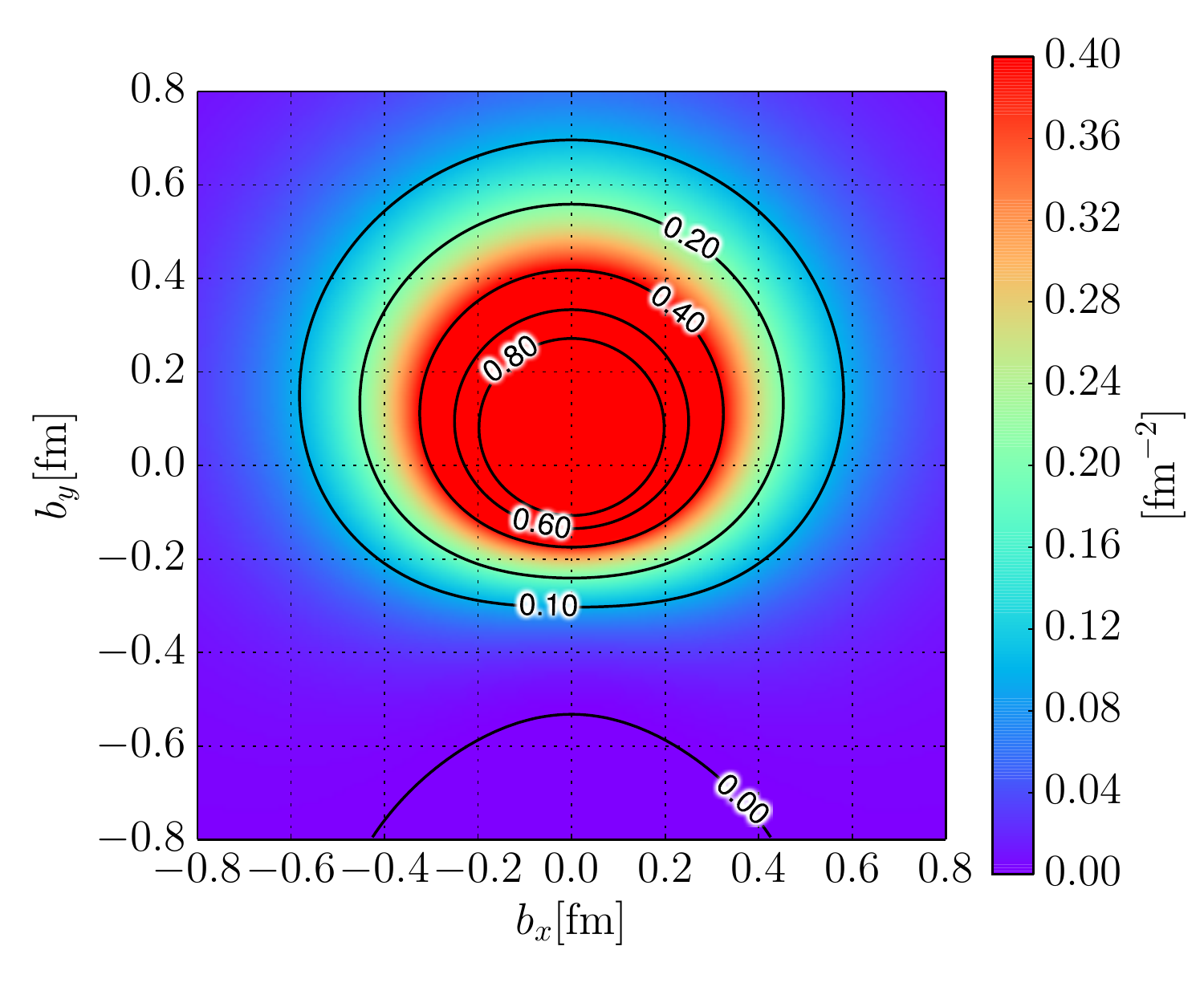}
\end{tabular}
\caption{(Color online) Unpolarized (left) and polarized (right)
  transverse quark-spin densities (TQSD) for $K^0 \rightarrow \pi^-$
  in the transverse impact-parameter plane ($b_x$-$b_y$), being
  calculated at $\mu=0.6$ GeV. We take the quark spin polarization as
  $s_x=+1$.}
\label{fig:4}
\end{figure}
When quarks involved in the $K\to \pi$ transition are not polarized in
the transverse plane, the transverse quark spin density is
defined only in terms of $A_{1,0}^{K\pi}$:
$\rho_1^{K\pi}(b)=A_{1,0}^{K\pi}(b^2)/2$, which is just the same as
the transverse charge density apart from the factor $1/2$.
The left panel of Fig.~\ref{fig:4} draws this transverse spin
density of the unpolarized quark in the $K\to\pi$ transition. The
result shows that the transverse spin of the quarks are uniformly
distributed. Note that the density is singular at $b=0$, which is very
similar to the transverse quark spin densities of the pion and the
kaon~\cite{Nam:2010pt,Nam:2011yw}.
On the other hand, if one of the quarks is polarized
along the $b_x$ direction, that is, $\bm s_\perp =(\pm1,0)$, then the
transverse quark spin density in the $K\to \pi$ transition gets
shifted to the positive $b_y$ direction, as shown in the right panel
of Fig.~\ref{fig:4}. It is of great use to compute the average shift
of the density so that we may see how much the transverse quark spin
density is distorted by the quark polarization. One can define the
average shift of the density to the $b_y$ direction as follows:
\begin{equation}
\label{eq:SHIFT}
\langle b_y\rangle^{K\pi}=
\frac{\int d^2b\,b_y\,\rho^{K\pi}_1(b,s_\perp) }
{\int d^2b\,\rho^{K\pi}_1(b,s_\perp) }
= \frac{1}{2m_K}\frac{B^{K\pi}_{T\,1,0}(0)}{A^{K\pi}_{1,0}(0)}.
\end{equation}
We obtain the numerical value $\langle b_y \rangle^{K\pi} = 0.169$
fm, which can be compared with those of the pion and the kaon. The
average shift of the transverse quark spin density in the pion was
obtained to be $\langle b_y\rangle^{\pi}=0.152$ fm that was almost the
same as the lattice calculation $\langle b_y\rangle^{\pi}=0.151\pm
0.024$ fm~\cite{Nam:2010pt}, whereas those in the kaon turned out to
be $\langle b_y\rangle^{K,u}=0.168$ fm and $\langle
b_y\rangle^{K,s}=0.166$ fm for the up and down quark components in
Model I in Ref.\cite{Nam:2011yw}. Thus, we find that the transverse
quark spin density of the $K\to\pi$ transition shows the largest
shift in comparison with those in the pion and the kaon.

\begin{figure}[t]
\includegraphics[width=8.5cm]{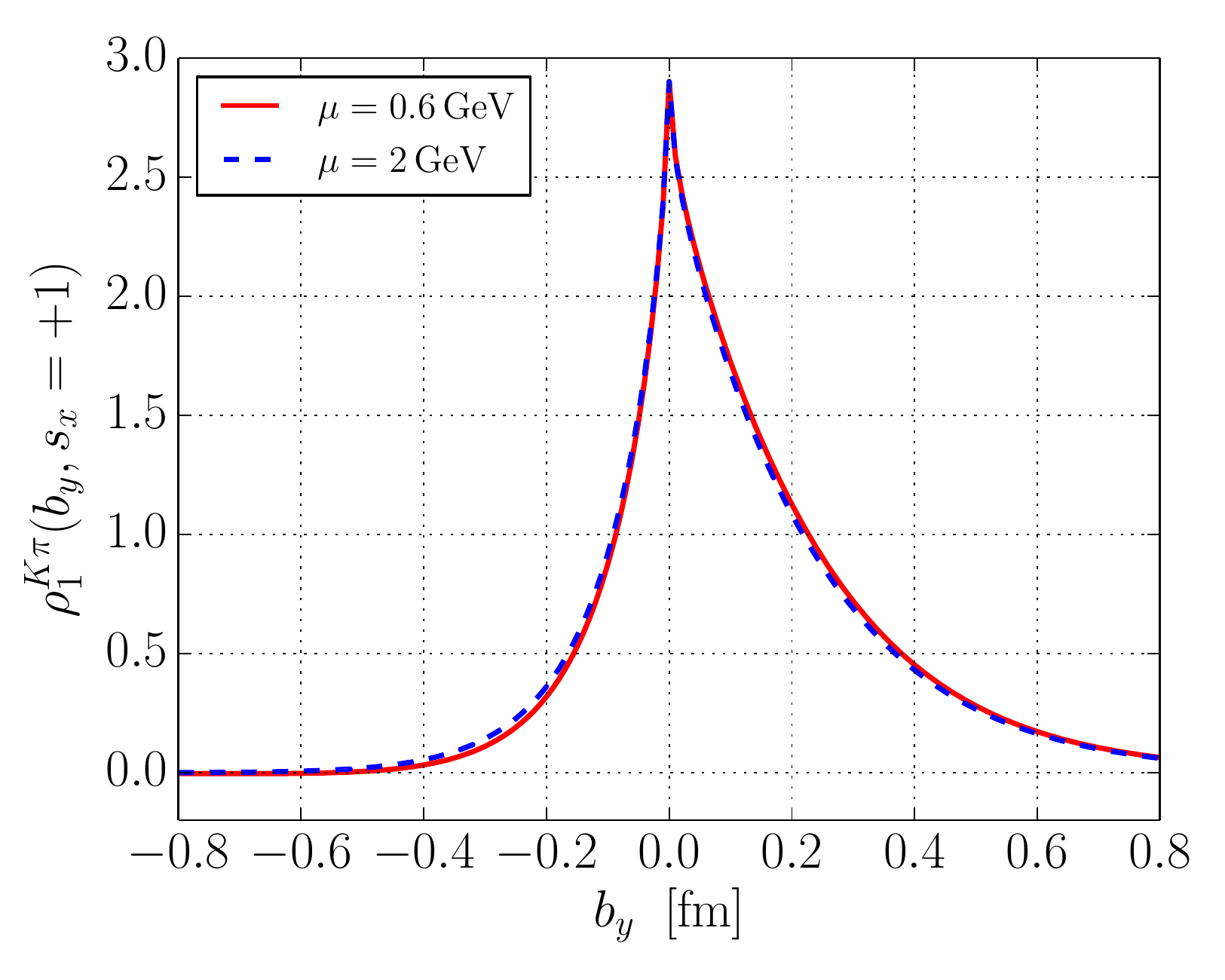}
\caption{(Color online) The profile of the polarized transverse quark
  spin densities at $\mu=0.6$ GeV (solid curve) and $\mu =2$ GeV
  (dashed curve), with $b_x=0$ fixed.}
\label{fig:5}
\end{figure}
Figure~\ref{fig:5} illustrates the profile of the polarized transverse
quark spin density of the $K\to \pi$ transition at two different
scales. It shows clearly the distortion of the density in the positive
$b_y$ direction. The scaling effect turns out to be negligible for the
transverse quark spin densities.

\section{Summary and conclusion}
In the present work, we have studied the transition vector and tensor
form factors for the $K\to \pi$ transition within the framework of the
nonlocal chiral-quark model from the instanton vacuum. We presented
the numerical results for the form factors and compared in particular
the tensor form factor with that of lattice QCD, considering the
renormalization group evolution. We also presented the results for the
transverse quark spin density in the course of the $K\to\pi$
transition without and with quark polarization in the transverse
direction.  We summarize below the important theoretical observations
in the this work:
\begin{itemize}
\item The vector and tensor form factors smoothly decease as $-t$
  increases. The value of the tensor form factor at $t=0$ becomes
  $B^{K\pi}_T(0)=0.792$ at the renormalization scale $\mu=0.6$ GeV and
  $B^{K\pi}_T(0)=0.709$ at $\mu=2.0$ GeV, while its overall $t$
  dependence does not change much.
\item The transition vector and tensor form factors can be
  parameterized by a $p$-pole type one, which is a
  function of $M$ and $p$, resulting in $(p,M) \approx(2.172,
  0.776\,\mathrm{GeV})$.
\item The present theoretical result
  $f^{K\pi}_{\mathrm{NL}\chi\mathrm{QM}}(0)=0.45$ for the transition
  tensor form factor at $t=0$ is in good agreement with that from
  lattice QCD $f^{K\pi}_\mathrm{LQCD}(0)=0.417\pm 0.014
  (\mathrm{stat})\pm 0.05
  (\mathrm{syst})$ at $\mu=2$ GeV.
\item The transverse quark spin density of the $K\to \pi$ transition
  was also computed as a function of the impact parameter $b$. When
  a quark in the course of $K\to \pi$ transition is polarized in the
  $b_x$ direction, the density becomes shifted to the positive $b_y$
  direction. The average shift of the density $\langle b_y
  \rangle^{K\pi} = 0.169$ fm at $\mu =0.6$ GeV is larger than those of
  the pion and the kaon.
\end{itemize}

In the present work, we wrote explicitly the expressions for the weak
transition generalized parton distributions that include all
information about the $K\to \pi$ transition. These generalized parton
distributions can be studied within the same theoretical
framework. The corresponding investigation is under way.

\section*{Acknowledgments}
H.Ch.K is grateful to Atsushi Hosaka for the discussions and his
hospitality during his visit to Research Center for Nuclear
Physics, Osaka University, where part of the present work was carried
out. He also wants to express his gratitude to Emiko Hiyama at RIKEN
for the valuable discussions. This work is supported by Basic Science
Research Program through the National  Research Foundation of Korea
(NRF) funded by the Korean government (MEST) (No. 2013S1A2A2035612
(H.D.S and H.Ch.K.)), respectively. The work of S.i.N. is supported in
part by the Korea Foundation for the Advancement of Science and
Creativity (KOFAC) grant funded by the Korea government (MEST)
(20142169990).


\end{document}